\documentstyle[11pt]{article}
\raggedright
\newcommand{\cB}{{\cal B}}
\newcommand{\cH}{{\cal H}}
\newcommand{\ei}[1]{{\scriptscriptstyle #1}}

\begin{document}
\title{Quantum channels\\
of the Einstein-Podolski-Rosen kind
\footnote{dedicated to Jan \L opusza\'nski, friend and colleague}}
\author{Armin.~Uhlmann}
\date{Institut f.~Theoretische Physik\\
Universit\"at Leipzig, Germany}
\maketitle
\begin{abstract}
An EPR-channel consists of two Hilbert spaces, $H^{\ei A}$ and $H^{\ei B}$,
and of a density operator sitting on the their direct product space
$H^{\ei A \ei B}$. The channel is triggered by a von Neumann measurement
on $H^{\ei A}$, resulting in a state (density operator) $\omega^{\ei A}$.
Because a measurement in $H^{\ei A}$ can be considered as a measurement
in $H^{\ei A \ei B}$ equally well, it induces a new state
in $H^{\ei A \ei B}$ and, hence, a new state, $\omega^{\ei B}$,
in $H^{\ei B}$. The map $\omega^{\ei A} \to \omega^{\ei B}$
depends only on the channel's original density operator, and not
on the chosen complete von Neumann measurement. This map is referred
to as ``channel map''.

The construction of the channel map is described together with
various of its properties, including an elementary link to
the modular conjugation and to some related questions.

The (noisy) quantum teleportation channel is treated as an example.
Its channel map can be decomposed into two EPR channel maps.
\end{abstract}

\section{Introduction}
In 1935 A.~Einstein, B.~Podolski, and N.~Rosen
posed an intelligent and far reaching question \cite{EPR35},
albeit suggesting a misleading answer. The abbreviation ``EPR''
in several of the following notations is pointing to these authors.
Further early contributions to the EPR-effect are due to
Schr\"odinger, \cite{Schr35a}, \cite{Schr35b}.
Since then a wealth of papers had appeared on the subject,
mostly discussing ``non-locality'' and similar aspects
which seem to contradict our causal feelings. They touch the
question whether and how space and time can live with the very
axioms of quantum physics, axioms which, possibly, are prior to
space and time. See \cite{Peres93} for a r\'esum\'e.

Quantum information theory considers the EPR-effect not as a
paradox but as a ``channel'', as (part of) a protocol to transfer
``quantum information'' from one system to another one
\cite{Ekert91}, \cite{BeWi92}.\\

The use of the word ``protocol'' may well be compared with
the way it is used in, say, ordinary telecommunication: It is
independent with what physical device the bits of its commands
are stored and processed.
The implementation independence in quantum information theory
is guarantied by the use of Hilbert spaces, states
(density operators), and operations between and on them. It
is {\it not} said, what they physically describe,
whether we are dealing with spins, polarizations,
energy levels, particle numbers, or whatever you can imagine.
Because of this, the elements of quantum information theory,
to which the EPR-channel belong, are of rather abstract nature.
Their physical realizations is generally much much more difficult
and often not visible yet.

The abstract setting of an EPR-channel starts with two Hilbert
spaces, $\cH^{\ei A}$ and $\cH^{\ei B}$, and their direct
product
\begin{equation} \label{ch1.1}
\cH^{\ei A} \otimes \cH^{\ei B} = \cH^{\ei A \ei B}
\end{equation}
and is completed with a density operator, $\varrho^{\ei A \ei B}$,
on it.

The subsystems, given by $\cH^{\ei A}$ and $\cH^{\ei B}$ respectively,
are referred to as A-- or B--system, or by its ``owners'',
Alice and Bob, who are responsible for local actions.
A local action of Alice is by definition a measurement
which can be performed by an observable of the A--system, or by
an operation which can be expressed by operators of the form
$X^{\ei A} \otimes 1^{\ei B}$, $X^{\ei A} \in \cB(\cH^{\ei A})$.
Similarly one defines Bob's local actions.

This definition of locality is compatible, though not equivalent,
with a possible localization of the
two subsystems in space. They may be sit even macroscopically
space-like one to another, but they must not do so.

Neither Alice nor Bob have access to all the operators (observables)
of the total system. Hence they can see the state
$\varrho^{\ei A \ei B}$ only partially. For instance, the state
$\varrho^{\ei A}$, induced in the A--system, is gained by  partial
tracing over the B--system, and is defined by
$$
{\rm Tr}_{\ei A} \varrho^{\ei A} X =
{\rm Tr}_{\ei A \ei B} \varrho^{\ei A \ei B}
( X \otimes 1^{\ei B} ), \quad \forall \, X \in \cB(\cH^{\ei A})
$$

A classical message is a sequence of letters from an alphabet or,
equivalently, a sequence of positions in a set which is
called ``alphabet''.
A quantum message is a sequence of positions in a state space of a
quantum system, hence a sequence of
states. A quantum channel is supposed to transfer the
quantum messages. It maps the state space of the sender
into that of the receiver.
Without some knowledge of the possible positions
used as letters, and of the channel's action, attempts to encode
the quantum message are hopeless. To be useful one needs some
additional classical message, transported through a classical
channel, and some conventions between sender and receiver.

The sender in a general EPR-channel
is a measuring apparatus in the A--system which performs a
von Neumann measurement, \cite{Neumann32}, \cite{Lued51}.
Let $X$ be the observable describing its action.
The duty of $X$
is to {\it prepare} one of the eigenstates of $X$, and
to {\it distinguish} it from the other ones by pointing to
its eigenvalue.
The physical meaning of the eigenvalues of $X$ is {\it not}
relevant for the purpose in question.

The state $\varrho^{\ei A \ei B}$ correlates the Alice' system
with that of Bob. These correlations constitute the channel.
Measuring $X$
destroys these correlations, thereby creating a new state in the
A--system and inducing a new one in Bob's system. Repeating this
procedure, which includes the regeneration of $\varrho^{\ei A \ei B}$,
Alice creates a random quantum message. The EPR-channel
transmits the quantum message to Bob, who receives, generally, a
deformed version of it. How much the message will be deformed
depends on the strength of the correlations provided by
$\varrho^{\ei A \ei B}$, but also on the choice of $X$ relative to
$\varrho^{\ei A}$.

Now we define the channel map. Let
\begin{equation} \label{Al.1}
\pi^{\ei A} = |\phi^{\ei A} \rangle\langle \phi^{\ei A} |, \quad
\phi^{\ei A} \in \cH^{\ei A}
\end{equation}
be a rank one projection operator. Let us assume $\pi^{\ei A}$ is a
non-degenerate eigenstate of $X$, and Alice's measuring apparatus
points to the eigenvalue associated with $\pi^{\ei A}$. (This happens with
probability
$\langle \phi^{\ei A}| \varrho^{\ei A} |\phi^{\ei A}\rangle$.)
Now, a measurement in a system is always a measurement in every
larger system, in our case in the AB--system. L\"uders' rule
provides us with the new state prepared by that measurement:
\begin{equation} \label{AB.1}
\varrho^{\ei A \ei B} \mapsto \omega^{\ei A \ei B} =
(\pi^{\ei A} \otimes 1^{\ei B}) \varrho^{\ei A \ei B}
(\pi^{\ei A} \otimes 1^{\ei B}) = \pi^{\ei A} \otimes \omega^{\ei B}
\end{equation}
Obviously, $\omega^{\ei B}$ is, up to normalization, the state
prepared in Bob's system by Alice' measurement.\\
The {\it channel map} is the map
\begin{equation} \label{AB.2}
\pi^{\ei A} \mapsto \omega^{\ei B} :=
\Phi^{\ei A \ei B}_{\varrho}(\pi^{\ei A}),
\quad \varrho = \varrho^{\ei A \ei B}
\end{equation}
We shall see that this map exists, i.~e.~it does not depend on
Alice's action.

\section{The maps $s^{\ei B \ei A}$ and $s^{\ei A \ei B}$ }
We start with EPR-channel maps {\it for vectors},
$\psi$, so that
$$
\varrho^{\ei A \ei B} = | \psi \rangle \langle \psi |, \quad
\psi \in \cH^{\ei A \ei B}.
$$

{\bf Lemma 1}

Let $\psi \in \cH^{\ei A \ei B}$ be written as a sum
\begin{equation} \label{sum1}
\psi = \sum \tilde \phi^{\ei A}_j \otimes \tilde \phi^{\ei B}_j
\end{equation}
with vectors $\tilde \phi^{\ei A}_i$ and $\tilde \phi^{\ei B}_k$
from $\cH^{\ei A}$ and $\cH^{\ei B}$ respectively. \hfill \\
Then the map
\begin{equation} \label{ch1.9}
\phi^{\ei A} \mapsto s^{\ei B \ei A} \, \phi^{\ei A},
\quad \phi^{\ei A} \in \cH^{\ei A},
\end{equation}
given by
\begin{equation} \label{ch1.10}
s^{\ei B \ei A} \, \phi^{\ei A}  = \sum_j
\langle \phi^{\ei A} | \tilde \phi^{\ei A}_j \rangle
\, \tilde \phi^{\ei B}_j ,
\end{equation}
is uniquely defined by $\psi$. Hence it can be denoted by
$s^{\ei B \ei A}_{\psi}$. \medskip

{\it Proof:} \, The idea is in assuming a von Neumann
measurement by Alice to check whether her system is in
the state given by $\phi^{\ei A}$.
If the answer is ``YES'', the vector
$$
\varphi := \{ \, | \phi^{\ei A} \rangle \langle \phi^{\ei A} |
\otimes 1^{\ei B}  \, \} \, \psi
$$
is prepared in the AB--system which can depend on $\phi^{\ei A}$
and $\psi$ only. By the help of (\ref{sum1}) this vector
is written
$$
\varphi = \phi^{\ei A} \otimes \sum
\langle \phi^{\ei A} | \tilde \phi^{\ei A}_j \rangle \tilde \phi^{\ei B}_j
$$
Comparing this expression with the definition (\ref{ch1.10})
we obtain the important relation
\begin{equation} \label{ch1.8}
\{ \, | \phi^{\ei A} \rangle \langle \phi^{\ei A} |
\otimes 1^{\ei B}  \, \} \, \psi =
\phi^{\ei A} \otimes s^{\ei B \ei A}_{\psi} \phi^{\ei A},
\quad \phi^{\ei A} \in \cH^{\ei A}
\end{equation}
This shows that the map (\ref{ch1.10}) does not depend on the
way the vector $\psi$ is represented as a sum (\ref{sum1}).
\medskip

{\bf Corollary 2} \, If Alice is successful in preparing $\phi^{\ei A}$
by a von Neumann measurement, the prepared vector
of the AB-System is given by (\ref{ch1.8}).  \hfill \\
\medskip

{\bf Corollary 3} \, Let $\phi^{\ei A}_1$, $\phi^{\ei A}_2, \dots$
be any collection of vectors satisfying
\begin{equation} \label{Al.2}
1^{\ei A} = \sum | \phi^{\ei A}_k \rangle\langle \phi^{\ei A}_k |
\end{equation}
then
\begin{equation} \label{Al.3}
\psi = \sum \phi^{\ei A}_k \otimes s^{\ei B \ei A}_{\psi} \phi^{\ei A}_k
\end{equation}

Indeed, this follows easily from (\ref{ch1.8}). The mathematical content
of lemma 1 is nothing than the well know theorem, stating that
the Hilbert space (\ref{ch1.1}) is isomorphic to the linear
Hilbert--Schmidt maps from $\cH^{\ei A}$ into the dual of
$\cH^{\ei B}$. One can map the latter by an antiunitary map
onto $\cH^{\ei B}$. This way we see, how quantum measurements
provide a randomly pointwise realization of Hilbert--Schmidt maps:\\
\medskip

{\bf Corollary 4} \, Every antilinear Hilbert--Schmidt map from
$\cH^{\ei A}$ into $\cH^{\ei B}$ can be identified with exactly one
of the maps $s^{\ei B \ei A}_{\psi}$.\\
\medskip

An interesting observation, \cite{Fivel95}, appendix\footnote{Thanks to
D.~DiVincenzo for the hint to Fivel's paper}, is the
{\it antilinearity} of (\ref{ch1.9}), clearly seen from its
definition (\ref{ch1.10}). Such a map cannot be tensored
with a linear one, for instance with the identity map of
another (complex!) Hilbert space. The sign
of the imaginary unit cannot be fixed in such a construct.
On the physical side this is very good: An antilinear map can
be represented by a linear one followed by time reversal.
A direct product of an antilinear and a linear map would
be equivalent, up to a linear operation, to reversing time in
the first but not in the other system, and this is forbidden.
Instead we have to apply (equivalents of) time reversal
simultaneously to {\it all} quantum systems which
can share entanglement. \hfill \\

A further important fact is the possibility to exchange
the roles of Alice and of Bob.
There is {\it no preferred direction} A $\to$ B
or B $\to$ A in the game, but a complete
symmetry with respect to the exchange A $\leftrightarrow$ B
in all equations and relations. \\
In particular the map
\begin{equation} \label{ch1.10a}
\phi^{\ei B} \mapsto
s^{\ei A \ei B}_{\psi} \, \phi^{\ei B}  = \sum_j
\langle \phi^{\ei B} | \tilde \phi^{\ei B}_j \rangle
\, \tilde \phi^{\ei A}_j, \quad \phi^{\ei B} \in \cH^{\ei B}
\end{equation}
is well defined, and there are counterparts to all the conclusions
above. In particular
\begin{equation} \label{ch1.8a}
\{ \, 1^{\ei A} \otimes
| \phi^{\ei B} \rangle \langle \phi^{\ei B} | \, \} \, \psi =
s^{\ei A \ei B}_{\psi} \phi^{\ei B} \otimes \phi^{\ei B},
\quad \phi^{\ei B} \in \cH^{\ei B}
\end{equation}
We mention some of the cross-relations between these maps.
With two arbitrary vectors $\psi$ and $\varphi$ from
$\cH^{\ei A \ei B}$ one has
\begin{equation} \label{trBA}
{\rm Tr}_{\ei A} s^{\ei A \ei B}_{\varphi} s^{\ei B \ei A}_{\psi}
=
{\rm Tr}_{\ei B} s^{\ei B \ei A}_{\varphi} s^{\ei A \ei B}_{\psi}
= \langle \psi, \varphi \rangle,
\end{equation}
To derive these equations, represents the vectors by any decompositions
$$
\psi = \sum \tilde \phi^{\ei A}_j \otimes \tilde \phi^{\ei B}_j,
\quad
\varphi = \sum \hat \phi^{\ei A}_j \otimes \hat \phi^{\ei B}_j
$$
Then
$$
s^{\ei B \ei A}_{\varphi} s^{\ei A \ei B}_{\psi} = \sum
| \tilde \phi^{\ei B}_j \rangle \langle \tilde \phi^{\ei A}_k,
| \hat \phi^{\ei A}_j \rangle \langle \hat \phi^{\ei B}_k |
$$
$$
s^{\ei A \ei B}_{\varphi} s^{\ei B \ei A}_{\psi} = \sum
| \hat \phi^{\ei A}_k \rangle \langle \tilde \phi^{\ei B}_j,
| \hat \phi^{\ei B}_k \rangle \langle \tilde \phi^{\ei A}_j |
$$
Taking the relevant trace one gets (\ref{trBA}). \\
The maps $s^{\ei B \ei A}$ and $s^{\ei A \ei B}$ are Hermitian
adjoints one from another. This is seen from the relation
\begin{equation} \label{Herm1}
\langle \phi^{\ei B} | s^{\ei B \ei A} \phi^{\ei A} \rangle
=
\langle \phi^{\ei A} | s^{\ei A \ei B} \phi^{\ei B} \rangle
\, \, \forall \, \phi^{\ei A} \in \cH^{\ei A},
\, \phi^{\ei B} \in \cH^{\ei B}
\end{equation}
which is shortly rewritten as
\begin{equation} \label{Herm2}
( \, s^{\ei B \ei A}_{\psi} \, )^* = ( \, s^{\ei A \ei B}_{\psi} \, )
\end{equation}

\section{EPR channel maps for states and observables}
Knowing a convenient description of EPR-channel maps for vectors,
we extend the formalism to Einstein-Podolski-Rosen
channels based on an arbitrary density operator. To start
with, the density operator, $\varrho^{\ei A \ei B}$, may be
in any decomposition
\begin{equation} \label{gen1}
\varrho^{\ei A \ei B} = \sum | \psi_i \rangle \langle \psi_i |,
\quad \psi_i \in \cH^{\ei A \ei B}
\end{equation}
According to lemma 1 and (\ref{ch1.10a}), every one of the
vectors $\psi_i$ gives rise to two antilinear maps
\begin{equation} \label{gen2}
\psi_i \leftrightarrow s_i^{\ei B \ei A}  \leftrightarrow
s_i^{\ei A \ei B}
\end{equation}
Again we ask for the state change if Alice' von Neumann measurement
confirms the state $|\phi^{\ei A}\rangle\langle\phi^{\ei A}|$
characterized by the vector $\phi^{\ei A}$. The
preparation causes the change
\begin{equation} \label{gen3}
\varrho^{\ei A \ei B}  \mapsto \Bigl(
|\phi^{\ei A}\rangle\langle\phi^{\ei A}| \otimes 1^{\ei B}
\Bigr) \, \varrho^{\ei A \ei B} \, \Bigl(
|\phi^{\ei A}\rangle\langle\phi^{\ei A}| \otimes 1^{\ei B}
\Bigr)
\end{equation}
There is no difficulty at all to insert the decomposition (\ref{gen1})
and to arrive, for all $i$, to a problem solved by lemma 1.
We use (\ref{ch1.8}) for the kets and, after respecting
(\ref{Herm1}, \ref{Herm2}), also for the bras. This way the
right hand side of (\ref{gen3}) is converted into
$$
(|\phi^{\ei A}\rangle\langle\phi^{\ei A}|)
\otimes \sum_i s_i^{\ei B \ei A} |\phi^{\ei A}\rangle
\langle\phi^{\ei A}| \, s_i^{\ei A \ei B}
$$
This expression depends only on $\varrho^{\ei A \ei B}$
and on $|\phi^{\ei A}\rangle\langle\phi^{\ei A}|$, the latter
dependence can be extended to arbitrary finite sums of positive
rank one operators and, for infinite dimensional Hilbert spaces,
to all trace-class operators.
By the same argument as in the proof of lemma 1 we
get, therefore,
\medskip

{\bf Lemma 5}

There is a channel map
\begin{equation} \label{gen5}
\omega^{\ei A} \mapsto \Phi_{\varrho}^{\ei B \ei A}( \omega^{\ei A} ),
\quad \varrho \equiv \varrho^{\ei A \ei B}
\end{equation}
such that for all
$$
\pi^{\ei A} = |\phi^{\ei A}\rangle\langle\phi^{\ei A}|, \quad
\phi^{\ei A} \in \cH^{\ei A}
$$
one gets
\begin{equation} \label{gen6}
\Bigl( \pi^{\ei A} \otimes 1^{\ei B}
\Bigr) \, \varrho^{\ei A \ei B} \, \Bigl(
\pi^{\ei A} \otimes 1^{\ei B} \Bigr)
= \pi^{\ei A} \otimes
\Phi_{\varrho}^{\ei B \ei A}(\pi^{\ei A})
\end{equation}
This defines a map map from the trace-class operators
on $\cH^{\ei A}$ into those of $\cH^{\ei B}$.
Every decomposition (\ref{gen1}) yields
\begin{equation} \label{gen7}
\omega^{\ei A} \mapsto
\Phi_{\varrho}^{\ei B \ei A}( \omega^{\ei A} ) =
\sum_i s_i^{\ei B \ei A} \omega^{\ei A} s_i^{\ei A \ei B}
\end{equation}
$\Phi_{\varrho}^{\ei B \ei A}$ is called the {\it EPR channel map
from Alice to Bob, based on} $\varrho$.
\medskip

We see the following:

a) To be a genuine channel map, the definition of $\Phi^{\ei B \ei A}$
should not depend on the actions of Alice. This is obviously
true.

b) The maps $\Phi^{\ei B \ei A}$, though {\it not} completely positive
themselves, become so after sandwiching them between antiunitaries.
Using for the latter a conjugation, we see that their action
on density operators is that of a complete copositive operator,
see \cite{Woron76}. Hence we may call the maps (\ref{gen7})
{\it completely $^*$-copositive}.

c) We may exchange the roles of Alice and Bob getting
$\Phi^{\ei B \ei A}$, the channel map from Bob to Alice.

d) There is a one-to-one correspondence
$$
\varrho  \Longleftrightarrow
\Phi_{\varrho}^{\ei B \ei A}  \Longleftrightarrow
\Phi_{\varrho}^{\ei A \ei B}, \quad \varrho \in \cH^{\ei A \ei B}.
$$

e) There is a virtual ``dictionary'' translating
any property of density operators of the AB--system into a
property of the associated channel map, and vice versa. \hfill \\
\medskip

The dual of a channel map as described above is a
map from Bob's observables (operators) into those of Alice.  Its duty
is to look at what is going on in Alice'system by transporting her
expectation values to Bob. (In case of infinitely many degrees of
freedom there are circumstances, where it
is advisable just to start by mapping observables.)

To construct the dual we need for every operator $Y$, acting
on $\cH^{\ei B}$, an operator $X$, acting on Alice' Hilbert space
with the following property:

If von von Neumann measurement of Alice results in a pure state
density operator, $\pi^{\ei A}$, and if $\Phi_{\varrho}^{\ei B \ei A}$
is the channel map introduced above, then
\begin{equation} \label{*channel1}
{\rm Tr}_{\ei A}  \pi^{\ei A} X = {\rm Tr}_{\ei B}
\Phi_{\varrho}^{\ei B \ei A}(\pi^{\ei A}) Y
\end{equation}
That is, according to (\ref{gen7}),
$$
\langle \phi^{\ei A}| X |\phi^{\ei A} \rangle =
\sum \langle s_i^{\ei B \ei A} \phi^{\ei A} | Y |
s_i^{\ei B \ei A} \phi^{\ei A} \rangle
$$
Let us denote by $\Phi_{\ei A \ei B}^{\varrho}$ the wanted operator,
$$
X = \Phi_{\ei A \ei B}(Y)^{\varrho}.
$$
From (\ref{Herm1} or \ref{Herm2}) one gets
\begin{equation} \label{*channel2}
\Phi_{\ei A \ei B}(Y)^{\varrho}
= \sum_i s_i^{\ei B \ei A} Y^* s_i^{\ei B \ei A}
= \sum_i \bigl( s_i^{\ei B \ei A} Y s_i^{\ei B \ei A} \Bigr)^*
\end{equation}
Again we see the antilinearity.
At the first instant one may think it irrelevant: Should we
not restrict ourselves to selfadjoint (Hermitian) oberservables?
However, such a practice is an oversimplification. Indeed, any
normal operator, $Y$,  $Y^*Y = YY^*$, is a perfect observable.
Its complex eigenvalues may be read off as points from a screen.
The crux with the antilinearity is that: A sequence of points,
appearing on Bob's screen, is an affine deformation of those
seen by Alice {\it together with a reflection on a certain line}.
The latter comes from the complex conjugation of the eigenvalues,
enforced by the Hermitian conjugation $Y \to Y^*$.
In other words, the orientations of Alice' and Bob's screen
are mutually opposite. The determinant of the mapping between
the screens, if not degenerated, is negative.

The antilinearity of the channel map
produces similar effects on geometric Berry phases. \hfill \\

\section{An excursion to quantum teleportation }
The teleportation protocol \cite{BBCJPW93} of Bennett, Brassard,
Crepeau, Josza, Peres, and Wootters,
needs a quantum and a classical information channel.
Here we are concerned with the quantum one and its channel maps.\\
Quantum teleportation lives on the direct product of three
Hilbert spaces,
\begin{equation} \label{tele1}
\cH^{\ei A} \otimes \cH^{\ei B} \otimes \cH^{\ei C} =
                          \cH^{\ei A \ei B \ei C}
\end{equation}
At first, as in the EPR case, one has to have control on the transfer
of vectors. The input of the teleportation channel
consists of an (unknown) unit vector to be teleported,
multiplied by an auxiliary one, the {\it ancilla},
carrying ``entanglement'',
\begin{equation} \label{tele2}
\psi^{\ei A \ei B \ei C} = \phi^{\ei A} \otimes \psi^{\ei B \ei C},
\quad
\phi^{\ei A} \in \cH^{\ei A}, \quad \psi^{\ei B \ei C} \in
\cH^{\ei B} \otimes \cH^{\ei C}
\end{equation}
The channel is triggered by a complete measurement of the
AB--system with respect to an orthonormal basis
\begin{equation} \label{tele3}
\psi_j^{\ei A \ei B} \in \cH^{\ei A} \otimes \cH^{\ei B}, \quad
j = 1, 2, \dots
\end{equation}
It seems natural to consider the maps
\begin{equation} \label{tele4}
s^{\ei C \ei B}, \quad s_j^{\ei B \ei A}, \, \, j = 1, 2, \dots
\end{equation}
where the first one is defined by lemma 1 with respect of
$\psi^{\ei B \ei C}$. The other ones are associated to the
members $\psi_j^{\ei A \ei B}$  of the basis (\ref{tele3})
accordingly.\\
It is tempting to compose these maps to transporters
$s^{\ei C \ei B} s_j^{\ei B \ei A}$ from
$\cH^{\ei A}$ to $\cH^{\ei C}$. And, indeed, that is the essence
of the quantum part of the famous teleportation protocol.
One needs no further
assumption on the nature of the vectors (\ref{tele2}) and
(\ref{tele3}) to run the protocol, though its effectiveness
(or failure) depends critical on them.
This is the content of the following lemma. \hfill \\
\medskip

{\bf Lemma 6}

If, with the notations above, the measurement device acting on
$\cH^{\ei A \ei B}$ is pointing to the i-th vector of (\ref{tele3}),
the teleportation provides in $\cH^{\ei C}$ the vector
\begin{equation} \label{tele5}
\phi_i^{\ei C} := t_i^{\ei C \ei A} \phi^{\ei A},
\quad
t_i^{\ei C \ei A} := s^{\ei C \ei B} s_i^{\ei B \ei A}
\end{equation}
{\it Proof:} \, In (\ref{tele1}) the vector
\begin{equation} \label{tmp}
\varphi_i = \Bigl( \, |\psi_i^{\ei A \ei B}  \rangle \langle
\psi_i^{\ei A \ei B}| \otimes 1^{\ei C} \Bigr)
\psi^{\ei A \ei B \ei C}
\end{equation}
is prepared by the measurement. Choosing in $\cH^{\ei B}$
an orthonormal basis
$\{ \phi_j^{\ei B} \}$ gives the opportunity to write
$$
\psi^{\ei B \ei C} = \sum \phi_j^{\ei B} \otimes
s^{\ei C \ei B} \phi_j^{\ei B}
$$
and hence
$$
\varphi_i = |\psi_i^{\ei A \ei B}  \rangle \langle \psi_i^{\ei A \ei B}|
\, \sum_j | \phi^{\ei A} \otimes \phi_j^{\ei B} \rangle \otimes
s^{\ei C \ei B} \phi_j^{\ei B}
$$
Now we choose in $\cH^{\ei A}$ an orthonormal basis
$\{ \phi_k^{\ei A} \}$
to resolve the scalar product in the last equation:
$$
\varphi_i = \psi_i^{\ei A \ei B} \otimes \sum_{jk}
\langle \phi_k^{\ei A} | \phi^{\ei A} \rangle
\langle s_i^{\ei B \ei A} \phi_k^{ \ei A} | \phi_j^{\ei B} \rangle
s^{\ei C \ei B} \phi_j^{\ei B}
$$
Using antilinearity,
$$
\varphi_i = \psi_i^{\ei A \ei B} \otimes s^{\ei C \ei B} \sum_k
\langle \phi^{\ei A} | \phi_k^{\ei A} \rangle \sum_j
\langle \phi_j^{\ei B} | s_i^{\ei B \ei A} \phi_k^{ \ei A}   \rangle
\phi_j^{\ei B}
$$
The summation over $j$ results in $s_i^{\ei B \ei A} \phi_k^{ \ei A}$.
Now, again by antilinearity, the sum over $k$ comes down to
$$
s_i^{\ei B \ei A} \sum_k \langle \phi_k^{\ei A} | \phi^{\ei A} \rangle
\phi_k^{ \ei A}
$$
Thus, we finally get the assertion of the lemma:
\begin{equation} \label{tele6}
\Bigl( \, |\psi_i^{\ei A \ei B}  \rangle \langle
\psi_i^{\ei A \ei B}| \otimes 1^{\ei C} \Bigr)
\psi^{\ei A \ei B \ei C} = \psi_i^{\ei A \ei B} \otimes
s^{\ei C \ei B} s_i^{\ei B \ei A} \phi^{\ei A}
\end{equation}
We have seen that the i-th teleportation channel map is composed
of two antilinear Hilbert-Schmidt maps. Hence $t_i^{\ei C \ei A}$
is {\it linear} and of {\it trace class}. In estimating its
magnitude by a norm, an adequate one is certainly the
{\it trace norm}
\begin{equation} \label{tele7}
\parallel t_i^{\ei C \ei A} \parallel_1 := {\rm Tr} \,
\sqrt{t_i^{\ei A \ei C} t_i^{\ei C \ei A}}, \quad
t_i^{\ei A \ei C} := (t_i^{\ei C \ei A})^*
\end{equation}
This norm depends as folllows on the reduced density operators
\begin{equation} \label{tele8}
\varrho_i^{\ei B} = {\rm Tr}_{\ei A} |\psi_i^{\ei A \ei B}\rangle
\langle \psi_i^{\ei A \ei B}|, \quad
\varrho^{\ei B} = {\rm Tr}_{\ei C} |\psi^{\ei B \ei C}\rangle
\langle \psi^{\ei B \ei C}|
\end{equation}

{\bf Lemma 7}

The trace norm $\parallel t_i^{\ei C \ei A} \parallel_1$ is
the square root of the transition probability (fidelity) between
$\varrho_i^{\ei B}$ and $\varrho^{\ei B}$.\\
\medskip

There are two more or less straightforward generalizations of
lemma 6. At first, we can convert the channel maps for vectors
to one for density operators (states). In doing so we consider
the preparation $|\varphi_i\rangle\langle\varphi_i|$ with
$\varphi_i$ from (\ref{tmp}). In this expression we vary the
vector $\phi^{\ei A}$ of (\ref{tele2}) and add them up.
This tells us how to teleport, through the i-th channel, an
arbitrary density operator, say $\omega^{\ei A}$, to the
C--system. The transport is done by
\begin{equation} \label{tele9}
\omega^{\ei A} \Longrightarrow
t_i^{\ei C \ei A} \omega^{\ei A}  t_i^{\ei A \ei C}
\end{equation}
Resolving the map according to lemma 6, and applying lemma 5, we can
replace the pure vector $\psi^{\ei B \ei C}$ of (\ref{tele1})
by an arbitrary density operator $\varrho^{\ei B \ei C}$. Then,
with such an arbitrary ancilla, the i-th teleportation channel
map reads
\begin{equation} \label{tele10}
\omega^{\ei A} \Longrightarrow  \Phi_{\varrho}^{\ei C \ei B} \Bigl( \,
s_i^{\ei B \ei A} \omega^{\ei A}  s_i^{\ei A \ei B} \, \Bigr)
\end{equation}
with $\varrho \equiv \varrho^{\ei B \ei C}$.

\section{Something more about EPR channel maps}
To prepare the next section, and for its own sake, we return to
section 2 and add some further relations. Let $\psi$ be a vector
from (\ref{ch1.1}) and $\varrho^{\ei A}$ and $\varrho^{\ei B}$
its partial traces. Then
\begin{equation} \label{ch1.12}
s^{\ei B \ei A} \, s^{\ei A \ei B} = \varrho^{\ei B},
\quad
s^{\ei A \ei B} \, s^{\ei B \ei A} = \varrho^{\ei A}
\end{equation}
From this one deduces the polar decompositions
\begin{eqnarray}
s^{\ei B \ei A} &=&  j^{\ei B \ei A} \sqrt{\varrho^{\ei A}}
= \sqrt{\varrho^{\ei B}}  j^{\ei B \ei A},
\nonumber\\
s^{\ei A \ei B} &=&  j^{\ei A \ei B} \sqrt{\varrho^{\ei B}}
= \sqrt{\varrho^{\ei A}}  j^{\ei A \ei B}
\label{ch1.14}
\end{eqnarray}
The partial antiunitaries
$j^{\ei A \ei B}_{\psi} \equiv j^{\ei A \ei B}$ and
$j^{\ei B \ei A}_{\psi} \equiv j^{\ei B \ei A}$ are
uniquely fixed by demanding one or both of the conditions
\begin{equation} \label{support1}
j^{\ei A \ei B} j^{\ei B \ei A} = {\rm support} \, \varrho^{\ei A},
\quad
j^{\ei B \ei A} j^{\ei A \ei B} = {\rm support} \, \varrho^{\ei B}
\end{equation}
$\psi$ is called {\it completely entangled} (with respect to Alice)
iff the support of $\varrho^{\ei A}$ is the identity map of
$\cH^{\ei A}$. This is equivalent with
calling $\varrho^{\ei A}$ {\it faithful}, or with calling
$\psi$ {\it separating} with respect to
$\cB(\cH^{\ei A}) \otimes 1^{\ei B}$.  If this occurs,
the dimension of $\cH^{\ei B}$ cannot be smaller than that of
$\cH^{\ei A}$. In case the dimensions are finite and equal, $\psi$
is completely entangled with respect to Alice iff it does so to Bob.
If $\dim \cH^{\ei A}$ is finite, $\psi$ can be {\it maximally entangled}
with respect to Alice. That means, $\varrho^{\ei A}$ is proportional to
$1^{\ei A}$. If both dimension are finite and equal, maximal
entanglement with respect to Alice implies the same to Bob, and
the reference to one of them is not necessary. Indeed, in calling
a vector of a bipartite system {\it maximally entangled}, one
supposes finiteness and equality of the dimensions by implication.\\

The equations above can be established by the help of a
Gram--Schmidt decomposition of $\psi$. Denote by $\phi_j^{\ei A}$
the vectors of a complete orthonormal system of eigenvectors
of $\varrho^{\ei A}$, and $p_j$ the corresponding eigenvectors.
Then there are orthonormal eigenvectors $\phi_j^{\ei B}$ such
that
\begin{equation} \label{GS0}
\psi = \sum \sqrt{p_j} \phi_j^{\ei A} \otimes \phi_j^{\ei B}
\end{equation}
One immediately infers from its definitions
\begin{equation} \label{GS1}
s_{\psi}^{\ei B \ei A} \phi^{\ei A}_j = \sqrt{p_j} \phi^{\ei B}_j,
\quad
s_{\psi}^{\ei A \ei B} \phi^{\ei B}_j = \sqrt{p_j} \phi^{\ei A}_j
\end{equation}
and the Gram-Schmidt form of the s-maps,
\begin{eqnarray}
s_{\psi}^{\ei B \ei A} \phi^{\ei A}  &=& \sum_1^m
\sqrt{p_j} \, \langle \phi^{\ei A} | \phi^{\ei A}_j \rangle
\, \phi^{\ei B}_j,
\nonumber\\
s_{\psi}^{\ei A \ei B} \phi^{\ei B}  &=& \sum_1^m
\sqrt{p_j} \, \langle \phi^{\ei B} | \phi^{\ei B}_j \rangle
\, \phi^{\ei A}_j
\label{GS3}
\end{eqnarray}
Some of the eigenvalues of $p_k$ may be zero, and then the $k$--th
term in the Gram-Schmidt representation
will vanish automatically. For the $j$-maps we had to exclude
them explicitly. Hence, assuming (\ref{GS0}), we should write
\begin{eqnarray}
j_{\psi}^{\ei B \ei A} \phi^{\ei A}  &=& \sum_{p_j \neq 0}
\langle \phi^{\ei A} | \phi^{\ei A}_j \rangle \, \phi^{\ei B},
\nonumber\\
j_{\psi}^{\ei A \ei B} \phi^{\ei B}  &=& \sum_{p_j \neq 0}
\langle \phi^{\ei B} | \phi^{\ei B}_j \rangle \, \phi^{\ei A}
\label{ch1.13}
\end{eqnarray}
{\it Remarks:}\\
(a) For any given partial antiunitary
map $j^{\ei B \ei A}$ there are vectors
$\psi \in \cH^{\ei A \ei B}$ such that
$j^{\ei B \ei A} = j_{\psi}^{\ei B \ei A}$.

(b) $j_{\psi}^{\ei B \ei A}$ is itself Hilbert-Schmidt
iff there are only finitely many terms in the sums (\ref{ch1.13}).
In other words: If and only if $s^{\ei B \ei A}$
is of finite rank, there is $\psi'$ such that
$j^{\ei B \ei A} = s^{\ei B \ei A}_{\psi'}$.

(c) Obviously, we need not care of finite rank and Hilbert-Schmidt
conditions in dealing with finite dimensional Hilbert spaces.

(d) $\psi$ is completely entangled with
respect to Alice iff $j^{\ei A \ei B}$ is an antiunitary
map from $\cH^{\ei A}$ into $\cH^{\ei B}$. \hfill \\
\medskip

Finally, let us mention the action of an operator of the form
$X^{\ei A} \otimes Y^{\ei B}$,
\begin{equation} \label{unitary1}
\varphi = ( X^{\ei A} \otimes Y^{\ei B} ) \, \psi
\end{equation}
A look at (\ref{ch1.10}) and (\ref{ch1.10a}) shows
\begin{equation} \label{unitary2}
s_{\varphi}^{\ei B \ei A} = Y^{\ei B}  s_{\psi}^{\ei B \ei A} (X^{\ei A})^*,
\quad
s_{\varphi}^{\ei A \ei B} = X^{\ei A}  s_{\psi}^{\ei A \ei B} (Y^{\ei B})^*
\end{equation}
We get from this and (\ref{ch1.14}) with unitary factors,
$U^{\ei A}$ and $U^{\ei B}$,
\begin{equation} \label{unitary3}
j_{\varphi}^{\ei B \ei A} = U^{\ei B}  j_{\psi}^{\ei B \ei A} (U^{\ei A})^*,
\quad
j_{\varphi}^{\ei A \ei B} = U^{\ei U}  j_{\psi}^{\ei A \ei B} (U^{\ei B})^*
\end{equation}
where now $\varphi = (U^{\ei A} \otimes U^{\ei B})\psi$.
Let us assume $\psi$ maximally entangled with respect to Alice and
to Bob. Then the $j$ are antiunitaries, and it follows
from (\ref{unitary3}) the implication
\begin{equation} \label{unitary4}
U^{\ei B} = j_{\psi}^{\ei B \ei A} U^{\ei A} j_{\psi}^{\ei A \ei B}
\quad \Longrightarrow (U^{\ei A} \otimes U^{\ei B}) \psi = \psi
\end{equation}
The unitaries
$U^{\ei A} \otimes j_{\psi}^{\ei B \ei A} U^{\ei A} j_{\psi}^{\ei A \ei B}$
form the {\it local stabilizer group of} $\psi$.

\section{Operator lifts to $\cH^{\ei A \ei B}$}
With a pair of maps, one from Alice to Bob and and one from Bob to
Alice, one can
compose an antilinear (linear) map in $\cH^{\ei A \ei B}$
provided both are antilinear (or both linear). Here we are concerned with
the antilinear case only.\\
Let us denote the ``twisting'' operation doing this by $\tilde \otimes$.
Requiring antilinearity, the maps are characterized completely by
their actions on product vectors:
\begin{eqnarray}
j_{\psi} \tilde \otimes j_{\psi} \,
(\phi^{\ei A} \otimes \phi^{\ei B}) &=&
j_{\psi}^{\ei A \ei B} \phi^{\ei B} \otimes
j_{\psi}^{\ei B \ei A} \phi^{\ei A}
\nonumber\\
j_{\psi} \tilde \otimes s_{\psi} \,
(\phi^{\ei A} \otimes \phi^{\ei B}) &=&
j_{\psi}^{\ei A \ei B} \phi^{\ei B} \otimes
s_{\psi}^{\ei B \ei A} \phi^{\ei A}
\nonumber\\
s_{\psi} \tilde \otimes j_{\psi} \,
(\phi^{\ei A} \otimes \phi^{\ei B}) &=&
s_{\psi}^{\ei A \ei B} \phi^{\ei B} \otimes
j_{\psi}^{\ei B \ei A} \phi^{\ei A}
\nonumber\\
s_{\psi} \tilde \otimes s_{\psi} \,
(\phi^{\ei A} \otimes \phi^{\ei B}) &=&
s_{\psi}^{\ei A \ei B} \phi^{\ei B} \otimes
s_{\psi}^{\ei B \ei A} \phi^{\ei A}
\label{modcon1}
\end{eqnarray}
The first of these operators is a standard one: It is the
{\it modular conjugation} of Tomita-Takesaki's theory,
$$
J_{\psi} = j_{\psi} \tilde \otimes j_{\psi}
$$
In this theory one considers $\cH^{\ei A \ei B}$ as representation
space of a $^*$--representation of $\cB(\cH^{\ei A}) \otimes 1^{\ei B}$,
(see, for example, \cite{Haag93}, sections III.2 and V.2).
The {\it modular operator}, $\Delta_{\psi}$, and the operator $S_{\psi}$,
$$
\Delta_{\psi} = \varrho^{\ei A} \otimes (\varrho^{\ei B})^{-1},
\quad S_{\psi} = J_{\psi} \sqrt{\Delta_{\psi}},
$$
if they exist,  satisfy
\begin{equation} \label{DS}
\sqrt{\Delta} (j \tilde \otimes s) = s \tilde \otimes j,
\quad
S \, (1^{\ei A} \otimes \sqrt{\varrho^{\ei B}}) = s \tilde \otimes j
\end{equation}
If we can rely on a Gram-Schmidt decomposition (\ref{GS0}), then
\begin{eqnarray}
J_{\psi} \phi_j^{\ei A} \otimes \phi_k^{\ei B} &=&
\phi_k^{\ei A} \otimes \phi_j^{\ei B}, \quad p_j p_k \neq 0
\nonumber\\
J_{\psi} \phi_j^{\ei A} \otimes \phi_k^{\ei B} &=& 0,
\quad p_j p_k = 0
\label{modcon2}
\end{eqnarray}
Similarly explicit expressions one obtains for the other operators.\\
\medskip
There is a lot more to say, including the discussion of examples.
But, hopefully, also this sketchy paper is of use.

\subsection*{Acknowledgement}
Part of this work was completed during the 1998 Elsag-Bailey --
I.S.I.~Foundation research meeting on quantum computation.
I thank the organizers of the XXI Max Born Symposium for
the privilege to speak in honour of Jan \L opusza\'nski. Thanks
to B.~Crell for discussions.

\end{document}